\documentclass{PoS}

\usepackage{multirow}
\usepackage{amsmath}

\newcommand{\juno}{\textsc{Juno}} 
\newcommand{\nue}{$\mathrm{\nu_{e}}$} 
\newcommand{\anue}{$\mathrm{\overline{\nu}_{e}}$}

\title{\juno{}: a General Purpose Experiment for Neutrino Physics}

\ShortTitle{JUNO: a General Purpose Experiment for Neutrino Physics}

\author{\speaker{Marco GRASSI}%
         \thanks{MG acknowledges support from the Chinese Academy of Sciences President's International Fellowship Initiative grant 2015PM007.}\\
        IHEP, Chinese Academy of Sciences, Beijing, P.R.China\\
        E-mail: \email{mgrassi@ihep.ac.cn}\\
        on behalf of the \juno{} Collaboration}

\abstract{\juno{} is a 20 kt Liquid Scintillator Antineutrino Detector currently under construction in the south of China.
This poster reviews \juno{}'s physics programme related to all neutrino sources but reactor antineutrinos, namely 
neutrinos from supernova burst, solar neutrinos and geoneutrinos. }

\FullConference{XXVII International Symposium on Lepton Photon Interactions at High Energies\\
		17-22 August 2015\\
		Ljubljana, Slovenia}

\begin{document}

\section{Introduction}
\juno{} is a Liquid Scintillator Antineutrino Detector (LAND) currently under construction in the south of China (Jiangmen city, Guangdong province).
Once completed, it will be the largest LAND ever built, consisting in a 20~kt target mass made of Linear Alkyl-Benzene (LAB) liquid scintillator (LS),
monitored by roughly 18000 twenty-inch high-QE photomultipliers (PMTs) providing a \textasciitilde80\% photo-coverage. 
Large photo-coverage and large QE are two pivotal parameters of the experiment, which allow an unprecedented 3\% energy resolution at 1~MeV.
The current conceptual design report~\cite{CDR} foresees the LS to be contained in an acrylic sphere 12~cm thick and 36~m wide. 
The whole detector is immersed in a cylindrical water pool, acting both as a moderator for the environmental radioactivity, and as 
a cherenkov detector to tag and veto cosmic muons.
%
%
%
The ultimate control and minimal impact of calorimetry systematics is of maximal importance
to achieve the aforementioned energy resolution.
For this reason, a novel LAND design was introduced, where a 
second layer of small PMTs is used to provide a second calorimetry handle with complementary systematic budget, 
allowing a combined, more precise and accurate energy scale definition. 
This calorimetry redundancy system is still under optimisation considerations in the context of \juno{} physics.
\textsc{Juno}'s main physics goal is to determine the neutrino mass hierarchy by detecting reactor $\overline{\nu}_e$ coming from 
two nuclear power plants both 53~km distant from the detector,
but here we focus on \textsc{Juno}'s physics programme relying on all neutrino sources but reactors. In particular,
 neutrinos from supernova (SN) burst, solar neutrinos, and geonutrinos. A complete review of \textsc{Juno}'s
physics goals can be found in~\cite{yellow_book}. 

\section{Neutrino Physics at JUNO}

\noindent \textsc{ \textbf{Supernova Burst Neutrinos}}\\
\indent A SN is a stellar explosion that briefly outshines an entire galaxy, 
radiating as much energy as the Sun or any ordinary star is expected 
to emit over its entire life span. During such explosion, 99\% of 
the gravitational binding energy of the newly formed neutron 
star is emitted in the form of $\nu$.
The observation of SN $\nu$ is expected to play a relevant role both in  
particle physics and astrophysics. Here we focus on the latter, where a SN signal might
help answering several fundamental questions, such as (\textsc{i}) what are the conditions
inside massive stars during their evolution?, (\textsc{ii})
what mechanism triggers the SN explosion?, (\textsc{iii})
are SN explosions responsible for the production of heavy chemical elements?, and
(\textsc{iv}) is the compact remnant a neutron star or a black hole?
Each of these questions would deserve a dedicated section, but because of the limited space
we consider (\textsc{i}) as a case study.

The Standard Stellar Evolution Model describes temperature and density of a star
as a function of time and distance from its centre. Optical observations usually provide 
benchmark data to test it, but optical observations have little power in constraining
the model of the star's innermost layers. Indeed, the star's high density results in optical photons 
propagating mainly via diffusion, hence loosing all the information about the stellar core.
On the contrary, $\nu$s interact weakly with stellar matter, and they represent a
powerful tool to probe the inner structure of the star.
In the case a star close to its collapse, the $\nu$ production is dominated by thermal processes
(mainly $e^{+}$-$e^{-}$ annihilating into $\nu$-$\overline{\nu}$ pairs). That is, the $\nu$
production rate, and the $\nu$ mean energy, both increase significantly with temperature. As a result, 
the last stages of the star's nuclear burning produce the most abundant $\nu$ signal 
(called pre-SN $\nu$), easier to detect and powerful in describing the stellar evolution.

Fig.~1 shows the simulated inverse beta decay (IBD) event rate in \juno{} for the nearest possible SN progenitor 
(the red supergiant Betelgeuse) whose mass is taken to be 20 solar masses (M$_{\odot}$) at a distance of 0.2~kpc. 
The sudden drop in the rate around 0.6 day before the SN explosion is ascribable to a drop 
of the core temperature, mostly due to the silicon depletion of the core itself.  
In the case of a SN explosion, \juno{}'s capability to precisely measure the position of such a dip 
could serve as a discriminator for different progenitor star masses. 
Moreover, the quick rise starting few hours prior to core collapse 
makes \juno{} an ultimate pre-warning system of SN explosion,
extremely valuable to the astrophysics community.

For a typical galactic SN at 10~kpc, there will be more than 5000 signal events solely from the IBD channel.
However, several other $\nu$ interactions contribute to the total event rate.
They differ in terms of total yield, energy spectrum and energy threshold.
Fig.~2 shows all of them together, where $(\mathrm{E_d})$
is the deposited visible energy in the detector, $(\mathrm{E^{th}})$ is the energy threshold of
each process, 
$(\nu\text{-p})$     are the neutral current interactions on protons, 
$(\nu\text{-e})$     are the elastic scatterings on electrons,
$(^{12}\text{C NC})$ are the neutral-current-mediated carbon excitations,
$(^{12}\text{N CC})$ are the charged current \nue{} interactions on $^{12}$C, and
$(^{12}\text{B CC})$ are the the same charged current interactions initiated by \anue{}.

%

\noindent\begin{tabular*}{\textwidth}{ p{0.45\textwidth} p{0.02\textwidth} p{0.45\textwidth}   }
\includegraphics[height=5cm]{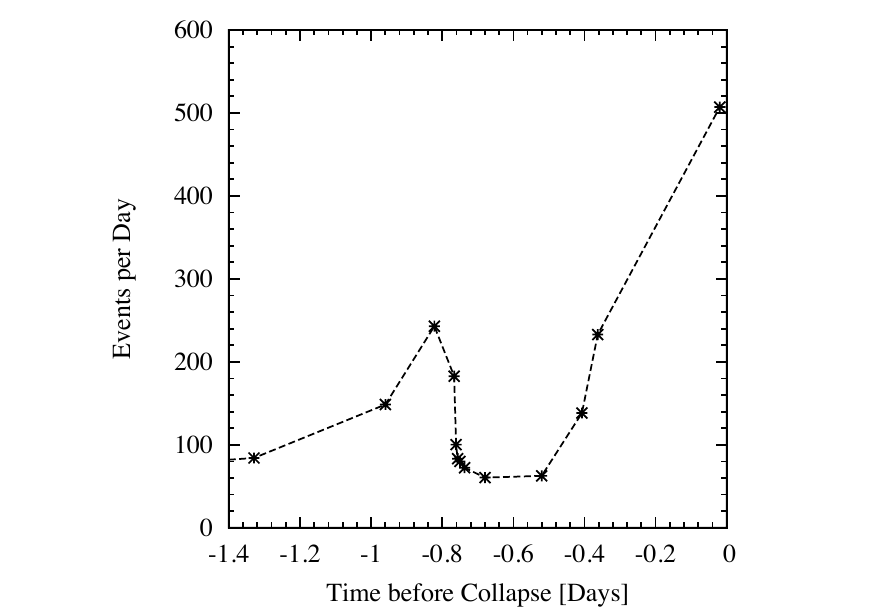}
&
&
\includegraphics[height=5cm]{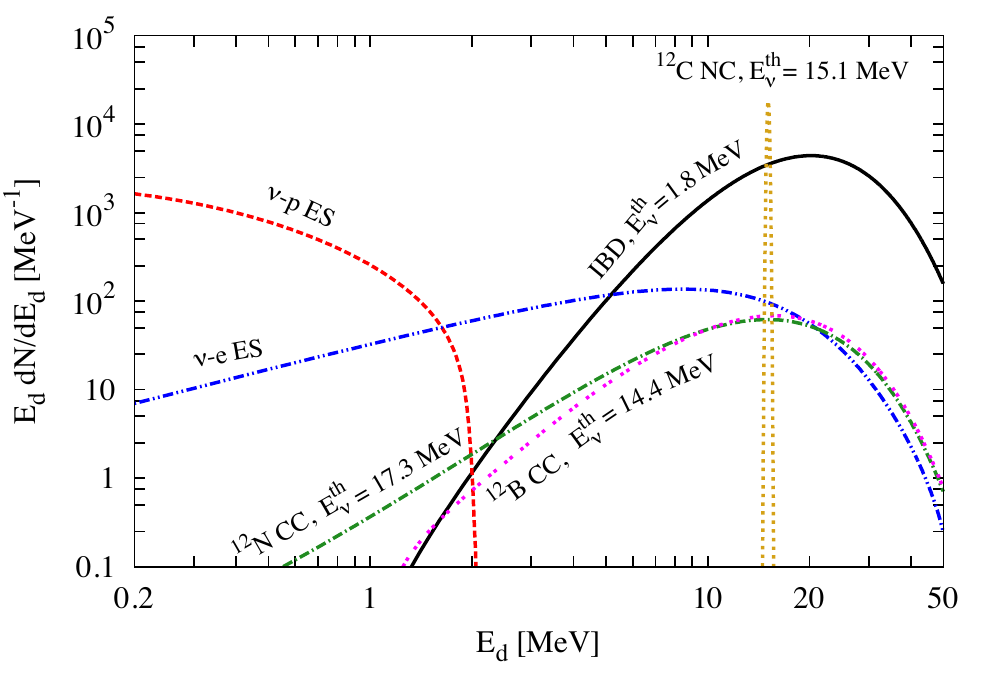} \\
{\small \textbf{Figure 1.} Neutrino event rate in JUNO for a massive star (20 
M$_{\odot}$) distant 0.2~kpc from the Earth, 
the same as the nearest possible SN progenitor Betelgeuse.\vspace{10pt}}
& &
{ \small \textbf{Figure 2.} Neutrino energy spectra 
in the \juno{} detector for a SN at 10~kpc, where no neutrino flavor conversions is assumed. 
$\mathrm{E_d}$ is the visible energy and $\mathrm{E^{th}}$ the threshold energy.
}\\
\end{tabular*}

\noindent \textsc{ \textbf{Solar Neutrinos}}\\
\indent The Sun is a powerful source of \nue{} with O(1~MeV) energy, produced in the thermonuclear 
fusion reactions happening in the solar core. \juno{}'s solar $\nu$  programme focuses on those emitted by the 
$^7$Be and $^8$B chains.
Indeed, despite the great achievements of the last decades, there are still important aspects of solar 
$\nu$ physics to clarify, and some questions of great relevance for astrophysics and elementary 
particle physics waiting for definite solutions. Two of the most important ones are
(\textsc{i}) the solution of the solar metallicity problem, 
and (\textsc{ii}) the detailed analysis of the 
oscillation-probability energy dependence in the lower end 
of the $^8$B $\nu$ spectrum.

(\textsc{i}) The solar metallicity problem emerged when
the former agreement between Standard Solar Model (SSM) and solar data 
got compromised by the revision of the solar surface heavy element content, 
leading to a discrepancy between the SSM and helioseismology results.
The predictions of different SSM versions differ (also) by the $^8$B and $^7$Be neutrino fluxes.
\juno{}'s capability to determine these fluxes with high accuracy,
together with data (coming from other future experiments) about the CNO fluxes, 
could help solving this key issue in nuclear astrophysics.

(\textsc{ii}) According to the Mikheyev-Smirnov-Wolfenstein (MSW) effect, $\nu$ oscillation
parameters are different if $\nu$ propagates through matter or in vacuum.
In the case of solar \nue{},
the transition between the two behaviors 
is expected to happen in the 1\textasciitilde3~MeV range, therefore
 solar $^8$B $\nu$s ---with their  continuous energy spectrum stretching far beyond 3~MeV ---
are a privileged tool to study the MSW-modulated energy dependence.
The theory predicts a smooth transition between 
the vacuum and matter related \nue{}-survival probabilities,
namely an up-turn in the spectrum. However,
none of the existing experiments so far observed a clear evidence of this effect.
The only exception is Super-Kamiokande, which got a mild evidence of the up-turn in its data~\cite{Renshaw:2013dzu}.
\juno{}'s capability to perform an independent and  high-significance 
test of the up-turn existence would be extremely important to confirm the consistency 
of the standard LMA-MSW solution, or to indicate any possible deviations from this standard paradigm.

\noindent\begin{tabular*}{\textwidth}{ p{0.45\textwidth} p{0.02\textwidth} p{0.45\textwidth}   }[t]
   \includegraphics[height=5cm]{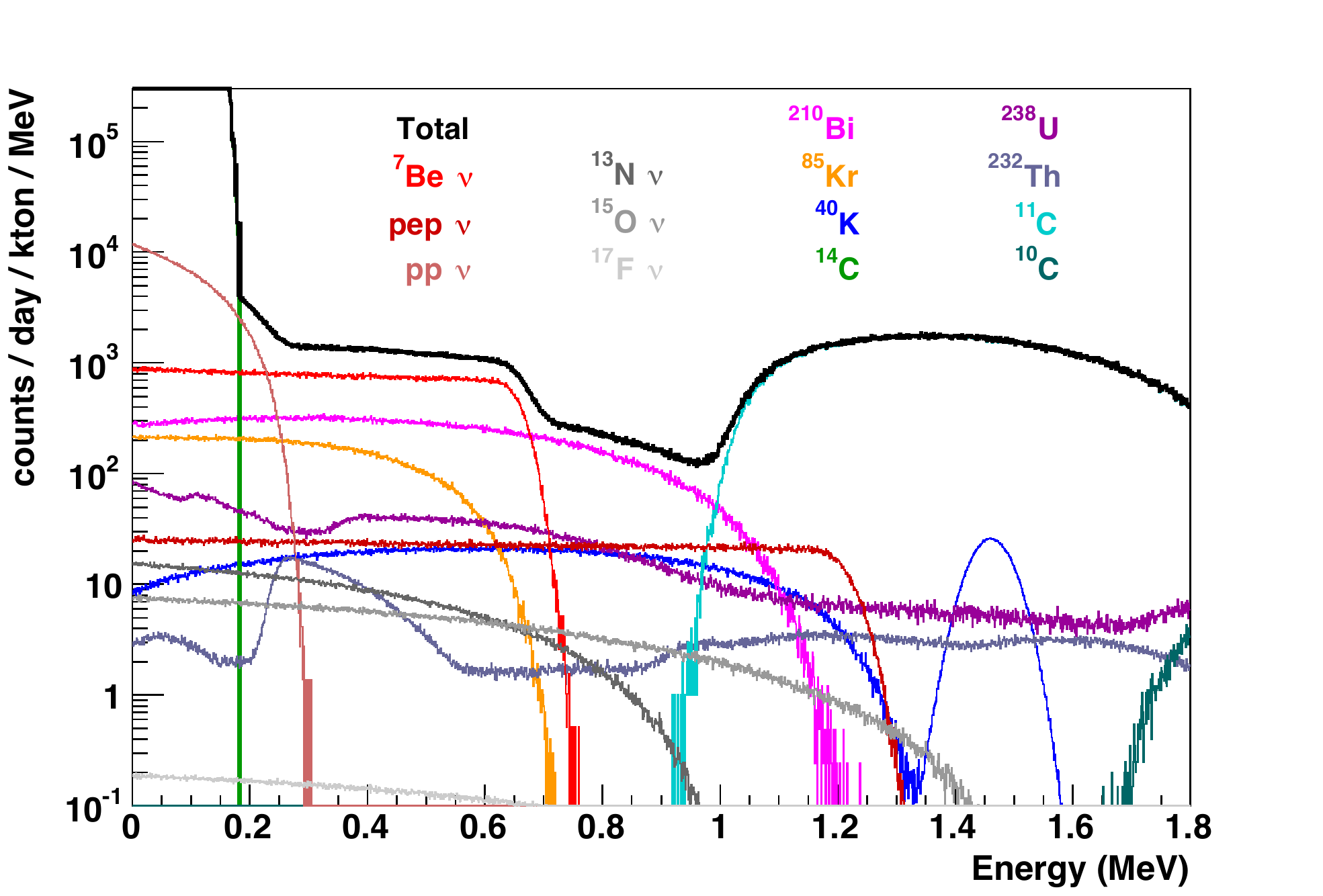}
&
&
\includegraphics[height=5cm]{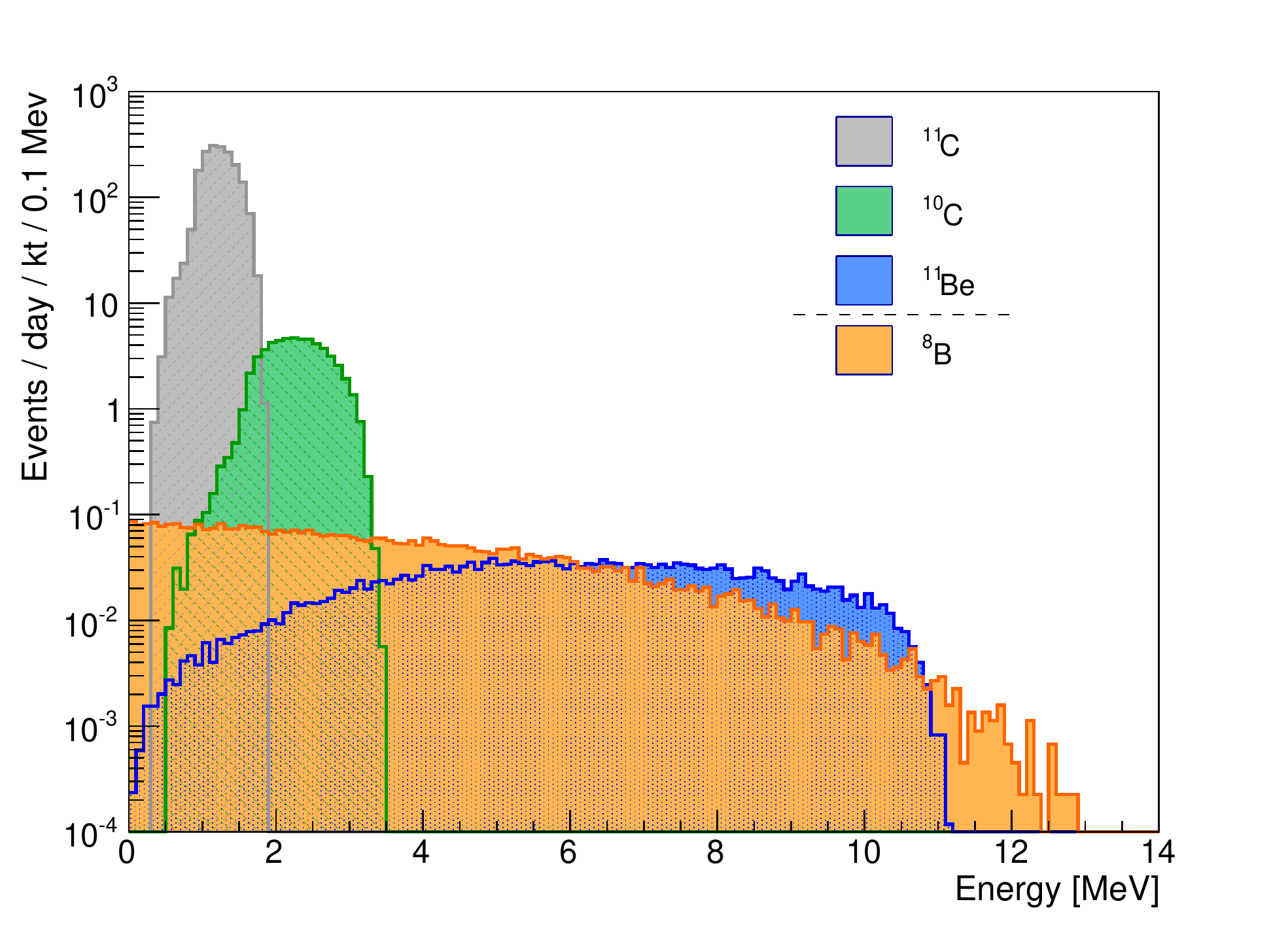}
\\
   {\small
   \textbf{Figure 3.} Energy spectra of singles from natural radioactivity (background) in the \textit{ideal}
   radiopurity scenario, together with the $^7$Be $\nu$ signal.\vspace{10pt}} 
& &
  { \small \textbf{Figure 4.} 
   Energy spectra of the $^8$B $\nu$ signal and of the cosmogenic backgrounds.}\\

\end{tabular*}

The challenge in detecting solar \nue{} at \juno{} is that they are detected only by elastic scattering,
which results in an experimental signature (single energy deposition) indistinguishable from most
of the background processes. The two main background sources are natural radioactivity and cosmogenic isotopes.
The first needs to be suppressed by achieving a high-level radiopurity in all the detector components. 
\textsc{Juno}'s \textit{baseline} radiopurity scheme envisages a $^{232}$Th, $^{40}$K, $^{14}$C residual contamination at
the level of $10^{-16}$, $10^{-16}$,  $10^{-17}$ g/g respectively. As a comparison, the same level of
radiopurity was achieved during KamLAND solar phase, and it would allow \juno{} to achieve a signal/background
ratio of 1/3. A more demanding radiopurity scheme (called \textit{ideal}) requires the previous contamination levels to improve
by one order of magnitude, which would correspond to Borexino phase I, and would allow a 2/1 signal/background ratio.
The energy spectrum of the radioactive background processes in the case of \textit{ideal} radiopurity, together with the signal 
$^7$Be \nue{}, are shown in Fig.~3.
Among the cosmogenic isotopes, the most dangerous are the long-lived ones, namely $^{10}$C ($\tau=24.4$~min) 
, $^{11}$C ($\tau = 27.8$~s), and $^{11}$Be ($\tau=19.9$~s) , since they cannot be suppressed by a muon veto
without introducing a relevant deadtime. The energy spectrum of these background events is shown in 
Fig.~4. The only way to handle them is to tag them via a three-fold coincidence (muon + spallation neutron
+ isotope decay) and subtract them statistically from the total spectrum.


\noindent \textsc{ \textbf{Geoeutrinos}}\\
\indent Over the last half a century, the Earth's surface heat flow has been established to be $46\pm3$~TW. 
However the community is still vigorously debating what fraction of this power comes from primordial 
versus radioactive sources. This debate touches on the composition of the Earth, the question of chemical 
layering in the mantle, the nature of mantle convection, the energy needed to drive plate tectonics, 
and the power source of the geodynamo, which powers the magnetosphere that shields the Earth from the harmful cosmic ray flux.
Radioactive beta-decays of heavy elements (such as Th and U) taking place inside the Earth 
result in an upwards \anue{} flux (also called geoneutrino flux) which can be detected at \juno{} by means
of IBD reactions. A precise measurement of such flux would allow to accurately define
the absolute abundance of Th and U in the Earth, which in turn would allow to:
(\textsc{i}) define the building blocks, the chondritic meteorites, that formed the Earth,
(\textsc{ii}) discriminate models of parameterised mantle convection that define the thermal evolution of the Earth,
(\textsc{iii}) potentially identify and characterize deep, hidden reservoirs in the mantle, and
(\textsc{iv}) fix the radiogenic contribution to the terrestrial heat flow.
Moreover, such studies can place stringent limits on the power of any natural nuclear reactor in or
near the Earth's core.

The main experimental challenge in detecting a geoneutrino signal is
to disentangle it from the reactor \anue{} signal, which is overwhelming. Such a separation can be done
only via statistical subtraction, and it relies heavily on a precise modeling of the 
low-energy reactor \anue{} spectrum. Moreover, to interpret the geoneutrino signal in terms of mantle's
radioactivity, the contribution from the Earth's crust need to be subtracted, 
since it's been well established that
the crust surrounding the detector will play a major role in total geoneutrino budget. 
Thus, to understand the relative contributions from the crust and mantle to the total 
geoneutrino signal at \juno{}, detailed geological, geochemical, and geophysical studies 
need to be performed in the areas surrounding the detector.

%
%
%

\section{Conclusion}
\textsc{Juno}'s physics programme is extremely broad, and makes it a genuine general purpose neutrino experiment. 
In this poster we presented only some of the topics associated to neutrinos not coming from nuclear reactor, 
namely supernova neutrinos, solar neutrinos and geoneutrinos. A complete review of \juno{}'s
physics programme can be found in~\cite{yellow_book}.
\vspace{4pt}

\noindent \textbf{\large Acknowledgments}\\
\indent MG acknowledges support from the CAS President's International Fellowship Initiative grant 2015PM007. 
MG wishes to thank Virginia Strati for helpful discussion on geoneutrinos.


\end{document}